# Measuring Voice UX Quantitatively: A Rapid Review

**KATIE SEABORN**

*Department of Industrial Engineering and Economics*
*Tokyo Institute of Technology*

**JACQUELINE URAKAMI**

*Department of Industrial Engineering and Economics*
*Tokyo Institute of Technology*

# Measuring Voice UX Quantitatively: A Rapid Review

**Katie Seaborn**
Tokyo Institute of Technology
Tokyo, Japan
seaborn.k.aa@m.titech.ac.jp

**Jacqueline Urakami**
Tokyo Institute of Technology
Tokyo, Japan
urakami.j.aa@m.titech.ac.jp

## ABSTRACT

Computer voice is experiencing a renaissance through the growing popularity of voice-based interfaces, agents, and environments. Yet, how to measure the user experience (UX) of voice-based systems remains an open and urgent question, especially given that their form factors and interaction styles tend to be non-visual, intangible, and often considered disembodied or "body-less." As a first step, we surveyed the ACM and IEEE literatures to determine which quantitative measures and measurements have been deemed important for voice UX. Our findings show that there is little consensus, even with similar situations and systems, as well as an overreliance on lab work and unvalidated scales. In response, we offer two high-level descriptive frameworks for guiding future research, developing standardized instruments, and informing ongoing review work. Our work highlights the current strengths and weaknesses of voice UX research and charts a path towards measuring voice UX in a more comprehensive way.

## CCS CONCEPTS

• Human-centered computing~Human computer interaction (HCI)  • Human-centered computing~Human computer interaction (HCI)~Interaction devices~Sound-based input / output

## 1 INTRODUCTION

Voice-based interfaces, agents, environments, and displays use natural and intuitive forms of interaction inspired by human-human communication. These technologies are not new, with several decades of history to be found in the field of human-computer interaction (HCI) and related disciplines [40,48]. Yet, modern advances in consumer devices and infrastructure, artificial intelligence (AI) and speech recognition, and natural language synthesizers have driven the widespread uptake of such systems in daily life. Smartphone-based voice assistants (VAs) like Apple's Siri and home smart speaker systems like Amazon's Echo with voice assistant Alexa are well-known consumer examples. Cutting-edge technologies like Google Duplex [34] are pushing the current standards to an extreme, where it is almost impossible to distinguish a computer voice from that of a real person. This recent and rapid progress has triggered a renewed interest in research on "the voice of the machine" [9,45]. In the CHI conference alone, there was a 2.46-fold increase in voice-related publications between 2015 and 2020.

The technical quality of voice recognition and voice expression is just one aspect of voice-based interaction. Decades of work on interactive voice response (IVR) systems, computer voice, conversational agents, and social robots has shown that user perception of voice plays a significant role [42]. People are affected by their pre-existing attitudes and behaviors towards computers but also tend to react to humanlike features of computer voice as they would react to people, without realizing it [31,40,42]. Moreover, there has been a growing recognition that social and psychological factors, including affect and emotion, trust, credibility, rapport, and the relational context, need to be considered [17,36]. This indicates a shift in focus from functionality and usability to user experience (UX).

The question of how to measure voice UX has subsequently been raised. A round table at CHI 2018 highlighted the need for understanding current practice as well as what approaches and methods should be used [22]. Clark et al. [9] reviewed the state of the art on speech user interfaces, covering research design, approaches to evaluation, and more. They noted that measurement work on speech interface UX is limited, in terms of where and with whom as well as with respect to validity and reliability. Seaborn et al. [45] noted similar issues for computer-based agents that use voice, calling for unification and standardization of quantitative measures. Kocaballi et al. [26] surveyed the conversational user interface (CUI) literature on measuring UX. They found a lack of consensus on what CUI UX is and how it is measured, leading them to develop a high-level unifying framework. Brüggemeier et al. [6] recognized a gap in measuring the UX of commercial voice assistants and then evaluated the validity of several instruments for Amazon Alexa. Nowacki et al. [43] proposed a set of heuristics for voice usability and UX, updating previous guidelines in light of changing technologies and conceptualizations of experience. Taken together, this body of work highlights raising interest in voice UX, as well as challenges that need to be addressed.

In continuation of this line of work, and with a view to answering the call made by the CHI 2018 panel, we pinpointed a gap at the intersections: measurement of voice UX. Kocaballi et al. [26] reviewed CUIs, which are not necessarily voice-based: other modalities, notably text, are used. Brüggemeier et al. [6] narrowly focused on voice assistants, excluding other "bodyless" voice-based possibilities. Nowacki et al. [43] focused on heuristics and guidelines rather than measurement. Clark et al. [9] and Seaborn et al. [45] each provided high-level reviews on approaches to the design and evaluation of speech interfaces—which may not have voice output—and computer agent voice—which may not have voice input. Here, we complement and extend this body of work by focusing on voice explicitly and comprehensively, especially audible and nonvisual, intangible form factors.

Underlying this work is also a recognition that UX itself is not easily defined. Indeed, lack of definitional and operational consensus has been a running problem [1,18,26,28,29]. Subsequently, despite the influx of voice UX research and a growing interest in measurement, there is yet no systematic overview of what is being done or how to do it. But such a review is necessary to achieve consensus and advance the field, as well as to develop reliable and valid measures, improve the quality of research and practice, and increase the comparability of studies.

In this paper, we provide an urgently needed state of the art to direct and support current and ongoing research. Using a gold standard rapid review process, we present a comprehensive description of how voice UX is being conceptualized and measured in HCI. We cover a variety of technologies that **rely on voice as the primary or sole means of interaction**, ranging from agents to interfaces to environments and displays. As yet, there does not seem to be consensus on a catch-all term for such systems. Here, we use **voice UX**, defining it as **verbal and auditory interaction with a computer-based system that uses voice-only input, voice-only output, and/or voice input-output**. As this is a late-breaking work, we restrict our focus to quantitative measures, which are common in HCI [50] and allow for ease of aggregation and comparison. We also restricted our databases to ACM and IEEE, major publishers of HCI work. We asked: **How is voice UX being measured quantitively?** We focused on three key sub-questions related to research design. These were: **RQ1. What factors are being experimentally manipulated as independent variables (IVs)? RQ2. What measures are being used to evaluate dependent variables (DVs?)** And **RQ3. What is the relationship between the IVs and DVs?** In this work, we are pre-emptively responding to the raising wave of studies on voice UX and increasing calls for an answer to the question of how to measure it. Our main contributions are threefold: (i) a description of the state of the art on quantitative voice UX measurement, (ii) an indication of emerging consensus and disagreement, as well as gaps and opportunities; and (iii) two frameworks for categorizing IVs and DVs in voice UX research that may be used to guide future research. This work acts a first step towards more comprehensive survey work that includes qualitative voice UX. We expect this work to help researchers in deciding which measures to choose and provoke initiatives on merging similar measures or developing new ones based on consensus.

## 2 METHODS

We conducted a rapid review of the literature following the Cochrane gold standard protocol [15]. Rapid reviews are used to generate a quick yet systematic descriptive and narrative summary of a focused area of study [12,14,23]. Importantly, they are conducted when a timely assessment is needed. As per the Cochrane protocol, we used the PRISMA checklist [35], adapted for HCI work[1], to structure or process and reporting. Our flow diagram is in the Supplementary Materials. Two researchers performed the review. The first author conducted the initial database searches and data extractions. Both authors then categorized the data, analyzed the data, and wrote up the results. The protocol for this study was registered (OSF # 423pq) before data extraction on December 10th, 2020[2].

---

[1] Some language and items related to the medical field (e.g., structured summaries) were modified or excluded.

## 2.1 Eligibility Criteria

We included full user studies involving at least one form of voice-only or primarily voice-based user interface, agent, display, environment, or some other interactive system. This included voice assistants, voice user interfaces, interactive voice response systems, conversational user interfaces, smart vehicles, and so on. We included studies that used at least one quantitative measure. Pilot studies, proposals, protocols, technical reports, literature reviews and surveys, and grey literature were not included. Papers were also excluded if there were not enough details to understand the measure and/or measurement. Only English papers were included.

## 2.2 Information Sources, Search Queries, and Study Selection

The ACM Digital Library and IEEE Xplore databases were queried on November 12th, 2020. All queries included the following keywords: "voice user interface," "voice assistant," "smart speaker," "interactive voice response." Given the amount of technical work in the IEEE database, index terms ("Publication Topics") were used in an attempt to filter results: "human computer interaction," "user experience," "human factors." The * qualifier was used to account for pluralization and different grammatical forms for each keyword. Manual additions from the paper citations and previous survey work were added. Papers were then screened independently by two raters for inclusion based on the eligibility criteria, each taking a random half of the set. A final 29 papers were included.

## 2.3 Data Collection, Items, and Analysis

Two researchers decided on the items for extraction based on the research question. They included the following, as stated in the pre-extraction protocol registration: description of studies (study type, research design, setting, agent type, agent device, participant demographics); IVs; DVs (measurements, subjective/objective, response format, validation). Each independently extracted metadata from half of the papers. The first author then generated descriptive statistics. Then, they conducted two inductive thematic analyses [5,16] to create categorical frameworks for each of the IV and DV data. First, they independently developed basic frameworks, then combined them through discussion, and then applied them to all 29 papers. Two rounds of coding were conducted for inter-rater reliability to be achieved, with Kappa values of .80+ per category. Only categories that achieved consensus were included. The second author then decided on the final attributions of themes where discrepancies in ratings existed.

## 3 RESULTS

Descriptive results on the study information are presented. Then, a visual timeline featuring the categories from the frameworks based on the thematic analyses of the IVs and DVs are presented. The tables for these are provided in their own sections with patterns highlighted and described. Finally, a matrix of the IV and DV categories is given.

### 3.1 Description of Studies

The surveyed papers reported on a variety of experiments and user studies (31 total). 66% (21) used a between-subjects design, and 34% (11) used within. 77% (24) were lab-based, 3 were in the field, and 4 were other (e.g., questionnaires). 59% (17) involved voice assistants, with the rest involving conversational agents (5), computer voice (2), in-car assistants (2), and three others (website, chatbot, and speech interface). Devices included smart speakers (38% or 11), computer speakers (17% or 5), an unknown speaker (2), smartphones (4), and one tablet. 1807 participants were involved across all studies: 667 men, 497 women, and two of another gender; a t-test did not find a significant difference between the number of men and women included across studies, $t = .82$, $p = .42$.

### 3.2 IVs and IV Framework

The framework resulting from the thematic analysis of the IVs is in Table 1. Relevant metadata and examples are provided along with descriptions for each category. Typical manipulations were varying the physical form of the voice system(s), vocal and social characteristics of the voice, conversation style, and interaction modality.

---
[2] https://osf.io/423pq

**Table 1: IV framework with context of study and examples.**

| Category and Description | Subcategory | Setting | Example | Sources |
|---|---|---|---|---|
| *Form Factor* | | | | |
| Manipulation of the form, style, and type of voice system | None | Lab, Field | Comparison of different devices; voice-only (disembodied systems) vs. embodied systems (e.g., robots) | [37]; [51]; [33]; [13]; [10]; [46] |
| *Voice Characteristics* | | | | |
| Manipulating sound as well as psychological characteristics of the voice output | Gender | Lab | Masculine vs. feminine voices | [10]; [30] |
| | Accent | Lab | Prestigious vs. regional accents; American vs. Swedish accents | [49]; [11] |
| | Emotions | Lab | Emotional expressions in graphical displays; emotionality of voice (energetic vs. subdued) | [47]; [41] |
| | Personality | Lab, Field | Personality type (e.g., extrovert vs. introvert); different characters (e.g., friendly vs. hostile) | [52]; [32]; [4] |
| *User Characteristics* | | | | |
| Comparing different groups of users or manipulating affective state of the user | Gender | Lab | Male and female participants | [30] |
| | Personality | Lab | Comparing extrovert/introvert participants | [32] |
| | Emotion | Lab | Driver emotion (happy/upset) | [41] |
| *Conversation* | | | | |
| Manipulating the voice system's response style, environment of use, or language style | Style | Lab | Verbal abuse style; response style; information style (e.g., low warning vs. high warning with details); conversational fillers; conversation style matching | [8]; [38]; [21]; [19] |
| | Context | Lab | Environment (e.g., living room vs. kitchen, vs. home office); baseline task; level of privacy; information content | [33]; [38]; [7]; [11] |
| | Natural | Lab | Simple commands; advanced instructions | [7] |
| *Interaction Modality* | | | | |
| Manipulating how users can interact with the voice system | Modality | Lab | Manipulating input modalities (e.g., voice vs. text vs. query form) | [39]; [20]; [53]; [47], [25] |
| | Query Form | Lab | Words vs. gaze activation | [39]; [20] |

Table 1 shows that each major category applies to several (3+) studies. Emotions and personality have received most attention for psychological factors. Manipulation of conversation and interaction modality were limited. While most are not unique to voice UX, one category stands out: Query Form. This refers to the use of a "wake word" or verbal query that triggers the start of an exchange with voice-based systems, especially smart speakers and voice assistants on smartphones, like Siri and Alexa. It applies to "bodyless" and ever-present voice systems, where there is not necessarily a visual or physical interface to begin or conduct interactions. The table also illustrates how most categories have relied on lab-based studies. Only Form Factor and Voice Characteristics involved field work.

Table 2: DV framework with measures/measurement grouped by type.

| Category | Subcategories | Type | Measure and/or Measurement |
|---|---|---|---|
| Usability | *None* | Obj | System task performance [39]; Exercise Behavior [37]; Gaze [16]; Language production (e.g., lexical complexity, adaptation) [51]; Interaction time [27]; Driving performance [38]; Disclosure (user responses) [53]; Accidents [41] |
| | | Subj | Continued use [33]; Satisfaction [33]; System effectiveness [24]; Tone clarity [8]; Satisfaction [3]; System Usability Scale (SUS) [6,44]; Speech User Interface Service Quality Questionnaire (SUISQ-R) [6]; Document feedback [38]; Voice understandability [10]; Perception of rapport [46]; Group decision performance [46]; Voice performance [11]; MeCue questionnaire [4] |
| Engagement | *None* | Obj | Interaction behavior (e.g., frequency and time) [37]; Emotional engagement (facial expressions) [47] |
| | | Subj | Cognitive engagement [47]; Time spent talking to virtual passenger [41]; Stimulation (UEQ modules) [4] |
| Cognition | Attention | Obj | Attentional allocation [27]; Situational awareness [38]; Driver attention [41] |
| | | Subj | Perceived attention [41] |
| | Workload | Obj | *None* |
| | | Subj | Mental workload [51]; User burden [13]; Driving activity load index [4] |
| Affect | *None* | Obj | Facial expressions (Affdex Software Development Kit – Affdex SDK) [4] |
| | | Subj | Negative impact [13]; User benefit [13]; Moral emotions (e.g., guilt) [8] Emotion manipulation check [41] |
| Sociality | *None* | Obj | Interaction behavior (e.g. gaze frequency, conversation turns) [27] |
| | | Subj | Social presence [32,33]; Intimacy [25,33]; Closeness (Subjective closeness index) [53]; Perceived power [46]; Conformity in decision-making [30]; Helpfulness [25]; Self-validation [25] |
| Attitudes | Trust | Obj | Trust (investment game) [49] |
| | | Subj | Trust [7,25,33]; Trustworthiness [30,46]; Safety [38] |
| | Likeability | Obj | *None* |
| | | Subj | Likeability [4,7,8,21,24]; Voice attractiveness [30]; Voice liking [11]; Information liking [11]; Attractiveness (UEQ modules) [4] Desirability (Marlowe-Crowne Social Desirability Scale - MCSDS) [10] |
| | Preference | Obj | *None* |
| | | Subj | Preference (self-reported) [33,38,46]; Smile-o-meter [52] |
| | Acceptance | Obj | *None* |
| | | Subj | Acceptance (self-reported) [7]; Acceptance Scale [4]; Familiarity [11] |
| Perception of Voice | Personality | Obj | *None* |
| | | Subj | Wiggins's interpersonal adjective scales [32]; Big Five Inventory [4,19] |
| | Anthropo-morphism | Obj | *None* |
| | | Subj | Anthropomorphism [8,46]; Human-likeness [21]; Voice gender [30] |
| | Intelligence | Obj | *None* |
| | | Subj | Perceived Intelligence [8,21,46] |
| Aggregate UX | *None* | Obj | *None* |
| | | Subj | Human-likeness [13]; Attractiveness differences [6]; Self-efficacy [7]; Interaction questionnaire [19]; Godspeed questionnaire [19] |

**Table 3: Matrix of the categories from the IV and DV frameworks. Circles indicate an intersection. Dots indicate no intersection.**

| | | Usability | Engagement | Cognition | | Affect | Sociality | Attitudes | | | | | | Perception of Voice | Aggregate UX |
| | | | | Attention | Workload | | | Trust | Likeability | Preference | Acceptance | Personality | Intelligence | Anthropo-morphism | | |
|---|---|---|---|---|---|---|---|---|---|---|---|---|---|---|---|---|
| **Independent Variables** | | | | | | | | | | | | | | | | |
| Form Factor | *None* | ○ | ○ | ○ | ○ | · | ○ | ○ | ○ | ○ | · | · | ○ | ○ | | ○ |
| Voice Characteristics | Gender | ○ | · | · | · | · | ○ | ○ | ○ | · | · | · | · | ○ | | · |
| | Accent | · | · | · | · | · | · | ○ | ○ | · | · | · | · | · | | · |
| | Emotion | ○ | ○ | ○ | · | ○ | · | · | · | · | · | · | · | · | | · |
| | Personality | ○ | ○ | · | ○ | ○ | ○ | ○ | ○ | ○ | ○ | ○ | · | · | | · |
| User Characteristics | Gender | · | · | · | · | · | ○ | ○ | ○ | ○ | · | · | · | ○ | | · |
| | Personality | · | · | · | · | · | ○ | · | · | · | · | ○ | · | · | | · |
| | Emotion | ○ | ○ | ○ | · | ○ | · | · | · | · | · | · | · | · | | · |
| Conversation | Style | ○ | · | ○ | · | ○ | · | ○ | ○ | ○ | · | ○ | ○ | ○ | | ○ |
| | Context | ○ | ○ | · | · | · | ○ | ○ | ○ | ○ | · | · | · | · | | ○ |
| | Natural | · | · | · | · | · | · | ○ | ○ | · | ○ | · | · | · | | ○ |
| Interaction Modality | Modality | ○ | ○ | · | · | ○ | ○ | ○ | · | · | · | · | · | · | | · |
| | Query Form | ○ | · | · | · | · | · | · | · | · | · | · | · | · | | · |

## 3.3 DVs and DV Framework

A variety of DVs were identified. Many studies used Usability measures, a traditional approach to measuring UX. Measurements of Cognition, such as Workload and Attention, also have a long tradition in ergonomics. Aside from these, we also found measurements of Affect, Sociality, and Engagement, signaling a shift to social factors. We also found instruments that measured several different aspects of experience together, which we call Aggregate UX.

Most measurements were self-report scales that used a Likert scale response format (49), although these varied from 5-point (35% or 17) to 7-point (39% or 19) to 9-point (3) to 10-point (20% or 10). 49% (43) were created by the researchers, 44% (39) were existing scales, and seven were modifications of these. 48% (42) were validated, while 37% were not, and 12 could not be determined. 14 papers reported validation; most (12) used Cronbach's alpha (M = .84, SD = .09), one used test-retest reliability (M = .85, SD = .09), and one used Spearman's rho (.78).

Table 2 presents the framework resulting from the thematic analysis of the DVs. A large number of measures and measurements fall under similar categories, notably usability (21). There are more subjective measures than objective (50 vs. 16); one category (i.e.., Aggregate UX) and several subcategories (e.g., Attitudes > Likeability, Cognition > Workload, Attitudes > Acceptance, etc.) have no objective measures. While most measures (as labelled) and measurements (as described) appear to have been stand-alone or unique, some were apparently used across several studies, including trust/trustworthiness (5 studies), likeability (5 studies), and preference (4 studies).

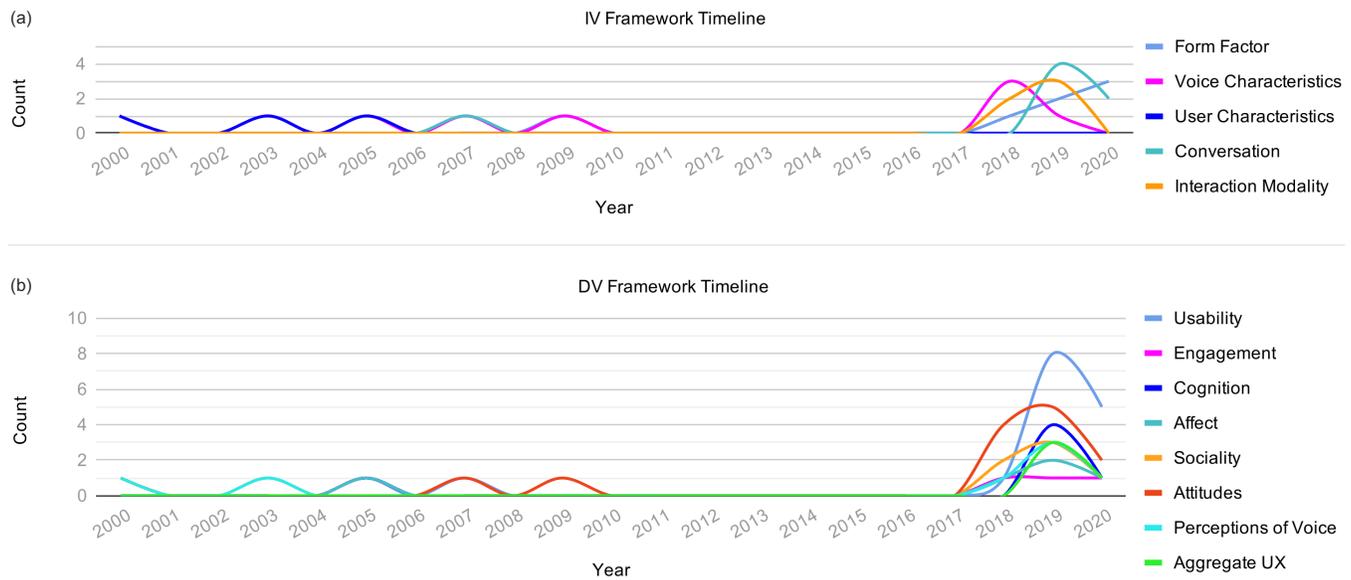

Figure 1: Timelines for (a) the IV framework and (b) the DV framework.

### 3.4 IV x DV Matrix

A matrix table was generated to compare, contrast, and contextualize the DV and IV frameworks against each other; see Table 3. The matrix shows that the DV categories of Usability, Engagement, Sociality, Trust, and Likeability are well-represented across the IV categories. Some gaps and intersections are notable. Cognition was not used for Conversation or Interaction Modality, even though it is generally considered relevant to these factors. Perceptions of Voice Intelligence was not manipulated or considered a factor in Conversation. Acceptance is a predictor for long-term use and should be examined. Aggregate UX, in contrast to Usability, has only been considered with respect to Form Factor and Conversation. However, Personality DVs and IVs overlap, indicating research design congruence.

Figure 1 presents two timelines for the IV and DV frameworks, showing how interest in categories have waxed and waned over time. Both indicate a peripatetic exploration of factors from 2000-2010, with a drop-off until a rapid increase starting in 2017. Figure 1a shows how Voice Characteristics was focused on earlier but has lost focus since 2010. Figure 1b shows a renewed interest in Usability and Attitudes, and a new interest in Aggregate UX.

## 4 DISCUSSION

Our rapid, systematic review of the voice UX literature on quantitative measures revealed a lack of consensus as well as several strengths and weaknesses. Many studies focused on the effects of "form factor" using usability measures. This represents the large variety of existing agent types and devices as well as a solid foundation on well-established usability work. An emerging focus on measuring psychological and social aspects shows that basic usability criteria might be satisfied by existing systems. Indeed, voice interaction is more than simple voice recognition; the field is starting to consider the context of use and importance of affective factors, trust, and sociality, as reflected in the recent panel [22] and related survey work [26]. Yet, there is a strong dependence on subjective measures based on self-reports. We need to develop objective measures (e.g., behavioral measures) to support these findings and validate subjective measurements. Additionally, while voice systems have a large range of applications, most studies were done in the lab. Relatedly, we did not find a single longitudinal study in our data set. Yet, we know that user expectation and attitudes can change with repeated use, and so this needs to be addressed. Overall, there was an extensive—we argue overly extensive—variety of measures for our DV categories. Our original goal was to present an overview of measures, but we had to give up on this idea because of this lack of agreement how to measure what. We now turn to what we can do about this state of affairs.

## 4.1 Next Steps and Future Work

This work marks a first step towards understanding the state of art on measuring voice UX and crafting a path forward. We offer four suggestions on what needs to be done next to address the gaps and improve the field:

- Start using the same measures and measurements for the same DVs, in lab and in field.
- Operationalize measures using theory about specific concepts (e.g., is rapport about trust?) and then merge measures that are highly related or equivalent.
- Create and validate a standardized aggregate UX tool for voice UX. Similar to how the Godspeed questionnaire for human-robot interaction was developed [2], use the IV and DV frameworks as a guide based on consensus.
- Develop objective measures where there are gaps as well as ones that complement subjective measures.

The next stage of this work will address its limitations by taking a comprehensive approach, including general databases (such as Scopus) and qualitative work. Future work will also summarize voice UX findings across studies.

## 5 CONCLUSION

We have systematically described the state of the art in quantitatively measuring voice UX. We have unearthed the strengths and weaknesses in this body of work, showing a lack of consensus. We urge researchers to consider our roadmap for improving the quality of voice UX research, especially in terms of standardization.

## ACKNOWLEDGMENTS

We thank the reviewers for their positive and useful feedback.